\documentclass[apjl,onecolumn]{openjournal}
\usepackage[T2A]{fontenc}
\usepackage[utf8]{inputenc}
\usepackage{amsmath,amssymb}
\usepackage{array}[=2016-10-06]
\usepackage{booktabs}
\usepackage{graphicx}
\usepackage{float}
\usepackage{tikz-cd}
\usepackage{ulem}
\usepackage{cancel}
\usepackage{orcidlink}
\usepackage{needspace}

\usepackage[dvipsnames]{xcolor}
\usepackage{xparse}
\definecolor{linkcolor}{rgb}{0.08, 0.38, 0.74}
\definecolor{citecolor}{rgb}{0.18, 0.55, 0.34}
\definecolor{urlcolor}{rgb}{0.03, 0.57, 0.82}

\usepackage{hyperref}
\hypersetup{
  colorlinks=true,citecolor=citecolor,linkcolor=linkcolor,urlcolor=urlcolor,
  pdftitle={Testing influence of  tides on gravitational wave emission with SPH code PHANTOM},
  pdfauthor={Marat Sh. Potashov; Sergei I. Blinnikov; Aleksandra V. Mishakina},
  pdfsubject={}
}

\newcommand{\code}[1]{\textsc{#1}}

\newcommand{\stkout}[1]{%
  \ifmmode
  \text{\sout{#1}}%
  \else
  \sout{#1}%
  \fi
}

\newif\ifshowtext
\showtexttrue 

\date{\today}

\shorttitle{Testing influence of  tides on GW emission with SPH code \code{phantom}}
\shortauthors{Potashov et al.}

\begin{document}

\title{Testing influence of  tides on gravitational wave emission with SPH code \code{phantom}\vspace{-3em}}

\author{Marat Sh. Potashov$^{\text{\color{blue}1,2,*}}$\orcidlink{0000-0002-0564-1101}}
\email[Marat Sh. Potashov (corresponding author): ]{marat.potashov@gmail.com}
\author{Sergei I. Blinnikov$^{\text{\color{blue}2,3}}$\orcidlink{0000-0002-5726-538X}}
\email[Sergei I. Blinnikov: ]{sblinnikov@gmail.com}
\author{Aleksandra V. Mishakina$^{\text{\color{blue}2,4}}$}

\affiliation{$^{\text{1}}$Keldysh Institute of Applied Mathematics, Russian Academy of Sciences, Moscow, Russia}
\affiliation{$^{\text{2}}$NRC Kurchatov Institute, Moscow, Russia}
\affiliation{$^{\text{3}}$Sternberg Astronomical Institute, Lomonosov Moscow State University, Moscow, Russia}
\affiliation{$^{\text{4}}$Moscow Institute of Physics and Technology, Moscow, Russia}

\begin{abstract}
  In this article, we demonstrate how a Newtonian SPH (Smoothed Particle Hydrodynamics) code \code{phantom}
  can be used to estimate the tidal contribution to gravitational wave (GW) waveforms.
  Using a binary neutron star (BNS) system with typical masses of 1.5 + 1.5 $M_\odot$ as an example,
  we show --- by comparing our results with waveforms calibrated against
  full numerical General Relativity (GR) simulations
  --- that the tidal contribution to GWs can be reliably
  extracted using as few as $\sim 10^4$ smoothed particles.
  Naturally, one must account for the difference in neutron star compactness between
  Newtonian gravity and GR.
  At separations greater than $\sim 80$ km, the relative tidal contribution falls below
  the level of numerical noise inherent in SPH.
  However, at such distances, GW waveforms for point-like sources are sufficient for detection purposes.
  It is also crucial to perform a hydrostatic/hydrodynamic relaxation of the initial models;
  otherwise, the entire GW signal is lost in the numerical noise.
\end{abstract}


\maketitle

\section{Introduction}

The problem of fast computing of GW-forms is still interesting for modern multimessenger astrophysics.
E.g., one of the most significant events was GW170817,
the first merger of the NS accompanied by a short gamma-ray burst GRB170817A~\citep{AbbottEtAl2017d}.
This event brilliantly confirmed the predictions of a short gamma-ray burst models made in the papers
\cite{BlinnikovEtAl1984a,BlinnikovEtAl1990},
see also more recent papers~\cite{BlinnikovEtAl2021,BlinnikovEtAl2022,KramarevEtAl2024,YudinEtAl2024},
however, alternative interpretations are feasible, and earlier discoveries are important.
The signal from GW170817 was detected by LIGO at $\sim 100$~s before GW170817.
Using fast GW-forms, one can detect the signal much earlier (at $\sim 10^3$~s) at LIGO-VIRGO-KAGRA
if it would be close enough.
One example of those fast fully analytical wave-forms is presented in our paper \cite{MishakinaBlinnikov2025a}
which improved the results of \cite{BuskirkHamilton2019}
who claimed fully analytical approach to the wave-form predictions
but actually relied on numerical integration of ODE.
The approach employed by \cite{MishakinaBlinnikov2025a} is applicable to point-like objects.
In reality one has to take into account tidal effects for finite-size bodies, see, e.g. \cite{ODellHamilton2024},
which we refer to as ``\code{Rising Tides}'' below.
Their model requires calibration against detailed GR numerical relativity (NR) results.

Direct NR simulations remain the most accurate reference.
The high computational cost of such calculations makes it
important to develop additional and less expensive methods
that can provide rapid preliminary estimates before applying NR.
SPH (Smoothed Particle Hydrodynamics) offers a possible alternative to grid-based hydrodynamics.
In a paper by \cite{MagnallEtAl2023}, which focuses on cosmological application
rather than the modeling of binary system mergers,
the authors report that particle-grid interpolation at each time step slows the evolution
relative to Newtonian $N$-body calculations.
They also note that applying their framework to compact-binary mergers
would require fixed mesh refinement.
Therefore, coupling SPH to a grid-based relativistic solver also incurs computational overhead.
Direct GW calculations with
\code{grphantom}~\citep[an SPH code with general-relativistic hydrodynamics,][]{LiptaiPrice2019}
likewise do not provide a fully self-consistent GR treatment.
The fluid evolves on a prescribed background metric
rather than in a spacetime evolved together with matter~\citep{ToscaniEtAl2021}.
In their TDE study, \cite{ToscaniEtAl2021} use a fixed Kerr metric, Newtonian self-gravity for the star,
and the weak-field quadrupole approximation for GW emission.
Thus, these approaches illustrate both the promise of SPH-based calculations
and the limitations that remain when modeling relativistic systems and their GW emission.

These considerations motivate us to test a hybrid candidate:
combining the analytical 6PN (6th-order post-Newtonian) point-mass orbital evolution
and waveform~\citep{MishakinaBlinnikov2025a}
with Newtonian SPH calculations that capture the finite sizes of the stars and their tidal response~\citep{PriceEtAl2018}.
In the present work, however, we restrict our analysis to the waveform amplitude.

Original \code{phantom} suffered from bugs that led to non-conservation of linear momentum,
as was observed by \cite{PotashovYudin2024}.
This is corrected for in the latest versions of \code{phantom} \citep{BernardEtAl2026}.
Currently, conservation of the angular momentum is also warranted,
which is very important for correct binary system evolution \citep{Marcello2017}.
These corrections allow us to apply the halted-pendulum relaxation (HPR) method,
following its description by \citep{KaltenbornEtAl2023}
and the recent results of Kudryashov, Potashov, and Yudin~(2026, in preparation),
to estimate the tidal response from quasi-equilibrium configurations rather than from a noisy SPH orbital evolution.

For the BNS (binary neutron star) system as an example,
we calculate quadrupole-based GW amplitudes at selected separations using these configurations
and compare the SPH tidal correction with the NR-calibrated \code{Rising Tides} model~\citep{ODellHamilton2024}.
We show waveform snapshots at selected fixed separations.
We identify the separation range in which the amplitude comparison is resolved above the SPH noise
and discuss the limitations of this amplitude-level test.
At fixed separation, we also apply a Love-number rescaling to the tidal contribution
to the strain amplitude to account for GR effects.
This amplitude correction does not provide a tidal correction to the orbital-frequency
evolution or accumulated GW phase.
Throughout this work, the orbital models and comparisons assume quasi-circular motion; eccentric binaries are outside our scope.

The outline of our present paper.
First, in Section~\ref{sec:compMBtoRT} we compare the approaches \cite{MishakinaBlinnikov2025a} vs. \code{Rising Tides} by \cite{ODellHamilton2024}.
After that, in Section~\ref{sec:phRes} based on the BNS case as an example,
the amplitude $h$ is compared across four approaches:
6PN (6th-order post-Newtonian, no tides),
\code{Rising Tides} approximation,
N (Newtonian, no tides),
and \code{phantom} (SPH with automatic account of tides).
Finally, we summarize our conclusions.

\section{Comparison of Analytical Methods: \code{Rising Tides} vs. Mishakina \& Blinnikov (2025)}
\label{sec:compMBtoRT}

We briefly compare the two analytical waveform approaches relevant to this work:
the point-mass 6PN model of \cite{MishakinaBlinnikov2025a}
and the NR-calibrated tidal model \code{Rising Tides} by \cite{ODellHamilton2024}.
The approaches share the goal of providing analytical gravitational-wave models, but differ fundamentally in scope, physical content, and mathematical strategy.
Table~\ref{tab:comparison} summarizes the key distinctions.

\begin{table}[h]
  \centering
  \caption{Methodological comparison of the two analytical approaches.}
  \label{tab:comparison}
  \begin{tabular}{p{3.2cm}p{5.2cm}p{5.2cm}}
    \toprule
    \textbf{Aspect} & \textbf{
    {\cite{ODellHamilton2024}}} & \textbf{
    \cite{MishakinaBlinnikov2025a}} \\
    \midrule
    Primary target & Binary neutron star (BNS) mergers, with tidal effects as the central physical ingredient & Neutron star and/or black hole binaries, treating them essentially as point masses during early inspiral  \\
    Physical effects modeled & Post-Newtonian inspiral + Backwards-one-Body merger + NRTidal phase/amplitude corrections, subsequently recalibrated against numerical relativity & Purely post-Newtonian inspiral in the point-mass approximation; no tidal deformability or finite-size effects are included \\
    Baseline waveform construction & Hybrid: PN inspiral + Backwards-one-Body merger; tidal terms added on top  & Direct analytical integration of the energy-loss equation within the PN framework  \\
    Nature of ``analytical'' result & Semi-analytical: tidal coefficients initially taken from NRTidal, then \textit{recalibrated} using NR data to improve merger-phase fits  & Fully analytical: closed-form expressions obtained via computer algebra (Maxima/Mathematica) without numerical interpolation of the waveform itself  \\
    Treatment of merger phase & Explicitly modeled via Backwards-one-Body and extended phase modeling beyond merger after recalibration  & Not modeled; the method is restricted to the \textit{earliest stages} of the inspiral, before merger  \\
    Validation strategy & Comparison against NR simulations (two equations of state); mismatch identified and then reduced by lifting coefficient constraints  & Comparison against numerical solutions; accuracy and limits of applicability quantified as functions of component masses  \\
    Intended application & High-accuracy waveform modeling for parameter estimation of BNS mergers, including tidal deformability extraction  & Rapid, low-latency template generation for early detection and multi-messenger alert triggering  \\
    Computational implementation & Python code \code{RisingTides}, steered by a Jupyter Notebook  & Computer algebra systems (Maxima, Wolfram Mathematica) \\
    Key limitation & Requires NR calibration for merger-phase accuracy; tidal model initially mismatches NR near merger  & Valid only for early inspiral on circular orbits; no merger, ringdown, or tidal physics included  \\
    \bottomrule
  \end{tabular}
  \vspace*{1.5em}
\end{table}

The two models also differ in how they are constructed.
The \code{Rising Tides} waveform combines a baseline binary-black-hole waveform, with PN inspiral matched to the Backwards-one-Body merger model and NRTidal phase, and amplitude corrections.
Its tidal coefficients, initially assumed to be independent of tidal deformability, are then recalibrated using NR data.
The model is therefore analytical in form but NR-calibrated in its coefficients;
this recalibration improves the merger-phase agreement and allows the phase model to extend beyond merger.

By contrast, Mishakina \& Blinnikov start from
\[
  \frac{dE}{dt} = -\mathcal{F}(v),
\]
with the flux expanded in PN orders, and derive $x(t) = v^2/c^2$ and hence the waveform $h(x)$ in closed form. The expressions are obtained symbolically in Maxima and Mathematica, producing a fully analytical template that does not rely on interpolation from NR data. The paper also notes that the earlier claim by \cite{BuskirkHamilton2019} of a fully analytical solution relied on numerically solving differential equations; Mishakina \& Blinnikov remove that step.

The approaches therefore serve different purposes. \code{Rising Tides} aims to model BNS merger waveforms, including tidal effects, and is suited to detailed source characterization. Mishakina \& Blinnikov prioritize computational speed and a fully analytical early-inspiral model, making it suited to early-warning detection, but it excludes matter effects and the merger phase.
For practical applications of their GW waveforms in LIGO--Virgo--KAGRA, they must be transformed to the time domain, as discussed in \cite{MishakinaBlinnikov2026}.

\needspace{6\baselineskip}
\section{The comparison of \code{phantom} simulations with other approaches}
\label{sec:phRes}

We now compare four waveform descriptions for an equal-mass binary neutron star (BNS):
the Newtonian point-mass model (N),
the relativistic 6PN point-mass model \citep{MishakinaBlinnikov2025a},
the NR-calibrated analytical \code{Rising Tides} (RT) model \citep{ODellHamilton2024},
and Newtonian SPH simulations that resolve the finite-size stars.

The BNS system has component masses $M_1=M_2=1.5\,M_\odot$ and total mass $M=3.0\,M_\odot$
(where $M=M_1+M_2$).
The symmetric mass ratio is $\eta=\mu/M=1/4$, with reduced mass
$\mu=M_1M_2/M$.
Each star is represented by a polytropic model with adiabatic index $\Gamma=2$
and radius $R_*=12$~km.
We define the star $A$ compactness as
\begin{equation}\label{eq:compactness}
  C_A=\frac{GM_A}{R_Ac^2}.
\end{equation}
For either star in this equal-mass system, $C_A\simeq0.1846$.
We consider quasi-circular orbits, neglecting eccentricity.
Throughout this section, we normalize the strain to the adopted distance to
Andromeda galaxy M31, $D_{\mathrm{M31}} \approx 0.78$~Mpc.

\subsection{Waveform polarizations and time-domain comparison}

We first define the Newtonian point-mass waveforms used as a reference.
In the leading-order quadrupole approximation, appropriate for weak gravitational
fields and slow source motion, the gravitational-wave strain at an observer a
distance $D$ from the source is obtained from the second time derivative of the
source mass quadrupole~\citep{bookLandauLifshitz1980,bookMisnerEtAl2017,ToscaniEtAl2021}:
\begin{equation}\label{eq:quadrupole-strain}
  h_{ij}^{\mathrm{TT}}(t;D)
  = \frac{2G}{c^4D}\,\Lambda_{ij,kl}\,\ddot{Q}_{kl}^{\mathrm{sys}}(t),
  \qquad
  Q_{kl}^{\mathrm{sys}}(t)=\int \rho(t,\mathbf{x})
  \left(x_kx_l-\frac{1}{3}\delta_{kl}|\mathbf{x}|^2\right)\,\mathrm{d}^3x.
\end{equation}
Here $\rho$ is the mass density,
$Q_{kl}^{\mathrm{sys}}$ is the trace-free source mass quadrupole of the entire
binary about its centre of mass, and $\Lambda_{ij,kl}$ projects the radiation onto its transverse-traceless
(TT) components in the observer's direction.
For an observer on the orbital angular-momentum axis (a face-on view),
only two components of $h_{ij}^{\mathrm{TT}}$ are independent
\begin{equation}\label{eq:face-on-polarizations}
  h_+(t)=\frac{h_{xx}^{\mathrm{TT}}-h_{yy}^{\mathrm{TT}}}{2},
  \qquad
  h_\times(t)=h_{xy}^{\mathrm{TT}}.
\end{equation}
For two point masses on a circular orbit, let $r$ be their centre-to-centre
separation, $\Phi(t)$ their orbital phase, and $\Omega=\dot{\Phi}$ their
orbital angular frequency.
The quadrupole projection then gives the Newtonian reference waveforms
\begin{equation}\label{eq:newtonian-waveforms}
  h_+^{\mathrm{N}}(t)=h_{\mathrm{N}}(r)\cos(2\Phi),
  \qquad
  h_\times^{\mathrm{N}}(t)=h_{\mathrm{N}}(r)\sin(2\Phi),
  \qquad
  h_{\mathrm{N}}(r)=\frac{4G^2M^2\eta}{c^4Dr}.
\end{equation}
The overall signs depend on the polarization convention.
The factor of two in the phase arises because the leading quadrupole signal
repeats twice during each orbital revolution.
Using Newtonian Kepler motion, $\Omega^2=GM/r^3$, gives the leading-order amplitude $h_{\mathrm{N}}(r)$.

For the 6PN point-mass model, we use the plus and cross polarizations given by Eqs.~(20) and (21), respectively,
in~\cite{MishakinaBlinnikov2025a}.

To include finite-size effects in a relativistic waveform model,
we next use \code{Rising Tides},
an NR-calibrated model for finite-size BNSs
that includes tidal corrections to $h_+$ and $h_\times$~\citep{ODellHamilton2024}
(also see Section~\ref{sec:compMBtoRT}).
In the zero-tide limit, \code{Rising Tides} reduces to the point-mass waveform
of \cite{BuskirkHamilton2019}, which numerically agrees
with the 6PN waveform of \cite{MishakinaBlinnikov2025a} used here.

The tide-induced stellar quadrupole is conventionally parameterized by the
dimensionless Love number $k_2$ in the linear response
\citep{Hinderer2008}:
\begin{equation}\label{eq:tidal-love-response}
  Q_{Aij}^{\mathrm{tide}}=-\lambda_A\mathcal{E}_{Aij},
  \qquad
  \mathcal{E}_{Aij}=\partial_i\partial_j\Phi_B^{\mathrm{ext}},
  \qquad
  \lambda_A=\frac{2}{3}k_{2,A}\frac{R_A^5}{G},
\end{equation}
Here $\Phi_B^{\mathrm{ext}}$ is the Newtonian gravitational potential generated
by companion $B$, and $\mathcal{E}_{Aij}$ defines the leading quadrupolar tidal
term in the Taylor expansion of this potential about the centre of star $A$.
$Q_{Aij}^{\mathrm{tide}}$ is the trace-free quadrupole of star $A$ about
its own centre of mass, whereas $Q_{ij}^{\mathrm{sys}}$ is the source quadrupole
of the entire binary about the binary centre of mass.
In Eq.~\ref{eq:tidal-love-response}, the subscript $B$ refers to the companion star,
$R_A$ is the stellar radius,
and $\lambda_A$ is the dimensional tidal deformability.
The corresponding dimensionless deformability is $\Lambda_A=(2/3)k_{2,A}C_A^{-5}$, where
$C_A$ is the compactness defined in Eq.~\ref{eq:compactness}.
For our $n=1$ ($\Gamma=2$) polytrope, the theoretical Newtonian value is
$k_2^{\mathrm{N}}=(15-\pi^2)/(2\pi^2)\simeq0.2599$
\citep{BrookerOlle1955}.
At the compactness of the neutron stars considered
here, $C\simeq0.1846$, the relativistic value is
$k_2^{\mathrm{GR}}\simeq0.0520$~\citep{Hinderer2008,Hinderer2009Erratum}.
For this equal-mass system, this gives $\tilde{\Lambda}\simeq161.8$ and,
in the convention of \code{Rising Tides}, $\kappa_{\mathrm{eff}}=
(3/16)\tilde{\Lambda}\simeq30.3$~\citep{ODellHamilton2024}.

In \code{Rising Tides}, $\kappa_{\mathrm{eff}}$ determines the tidal phase and
amplitude corrections.
The model applies the tidal correction to its
binary-black-hole (BBH) baseline strain, $h_{\mathrm{BBH}}$, to construct the
complex BNS strain~\citep{ODellHamilton2024}:
\begin{equation}\label{eq:rising-tides-strain}
  h_{\mathrm{RT}}(t)=h_{\mathrm{BBH}}(t)h_T(t),
  \qquad
  h_T(t)=A_T(t)e^{-2\mathrm{i}\phi_T(t)}.
\end{equation}
With the convention $h_{\mathrm{RT}}=h_+^{\mathrm{RT}}-\mathrm{i}h_\times^{\mathrm{RT}}$,
the polarizations are
\begin{equation}\label{eq:rising-tides-polarizations}
  h_+^{\mathrm{RT}}=\operatorname{Re}(h_{\mathrm{RT}}),
  \qquad
  h_\times^{\mathrm{RT}}=-\operatorname{Im}(h_{\mathrm{RT}}).
\end{equation}
The phase correction $\phi_T$ also changes the instantaneous GW frequency.
It is included in the time-domain \code{Rising Tides} curves shown below, but
we do not analyze this phase/frequency correction here.
Our quantitative comparison focuses on the strain-amplitude envelope at fixed separation.

Alongside this analytical model, we calculate the finite-size contribution
directly with the Newtonian smoothed particle hydrodynamics (SPH) code \code{phantom}.
Each neutron star is represented by a distribution of smoothed particles,
resolving the stars' finite extent and tidal deformation and their
contribution to the source quadrupole.
The SPH configurations considered below are corotating (tidally synchronized):
the spin angular velocity of each neutron star equals the orbital angular velocity,
$\Omega_{\mathrm{spin},1}=\Omega_{\mathrm{spin},2}=\Omega_{\mathrm{orb}}$.
At a fixed separation, such a configuration is quasi-stationary: its tidally
distorted shape is approximately time-independent in the frame rotating with the
binary.
We obtain these configurations with halted-pendulum relaxation (HPR),
which repeatedly sets the residual particle velocities in the rotating frame to
zero at maxima of the total corotating kinetic energy.
This suppresses residual
oscillations and lets the binary approach equilibrium~\citep{KaltenbornEtAl2023}.
In the BNS models used in this comparison,
the SPH resolution is $N=10^4$ particles per star.
Higher-resolution tests increased the HPR runtime but did not
substantially change the tidal response or its Love-number-rescaled comparison with
\code{Rising Tides}.

After the final HPR relaxation, we evolve each configuration at approximately
fixed separation for several GW cycles.
We discretize the source mass quadrupole similar to~\cite{ToscaniEtAl2021} but trace-free:
\begin{equation}\label{eq:sph-quadrupole}
  Q_{ij}^{\mathrm{sys}}=\sum_b m_b\left(x_{b,i}x_{b,j}
  -\frac{1}{3}\delta_{ij}|\mathbf{x}_b|^2\right),
\end{equation}
Here $m_b$ and $\mathbf{x}_b$ are the mass and position of SPH particle $b$,
with positions measured from the binary centre of mass.
The sum runs over particles in both stars,
so this is the SPH discretization of the total system quadrupole.
The corresponding SPH strain requires the second time derivative of this quadrupole.
For particles with inertial-frame velocities
$\mathbf{v}_b$ and accelerations $\mathbf{a}_b$, differentiating
Eq.~\ref{eq:sph-quadrupole} gives
\begin{equation}\label{eq:sph-quadrupole-second-derivative}
  \ddot{Q}_{ij}^{\mathrm{sys}}
  =\sum_b m_b\left[a_{b,i}x_{b,j}
    +2v_{b,i}v_{b,j}+x_{b,i}a_{b,j}
    -\frac{2}{3}\delta_{ij}\left(|\mathbf{v}_b|^2+
  \mathbf{x}_b\cdot\mathbf{a}_b\right)\right],
\end{equation}
following the particle-level expression in~\cite{ToscaniEtAl2021}.
In evaluating this expression for each relaxed binary configuration, we use
particle positions, velocities, and accelerations in the inertial (laboratory)
frame.
Substituting the resulting $\ddot{Q}_{ij}^{\mathrm{sys}}$ into
Eq.~\ref{eq:quadrupole-strain} gives $h_{ij}^{\mathrm{TT}}$; applying the
face-on projection in Eq.~\ref{eq:face-on-polarizations} then gives
$h_+^{\mathrm{SPH}}$ and $h_\times^{\mathrm{SPH}}$.
Because these Newtonian runs omit radiation-reaction-driven
inspiral, they provide quasi-stationary waveform samples, not a complete inspiral waveform.

We now compare the time-domain gravitational-wave polarizations,
$h_+$ and $h_\times$, predicted by the four descriptions considered here:
Newtonian point masses, 6PN point masses, 6PN with the \code{Rising Tides}
correction, and the SPH calculation.
The \code{Rising Tides} waveforms include both tidal amplitude
and phase corrections.
The latter also changes the instantaneous GW frequency.
For these time-domain comparisons, we set the common initial GW phase from the
SPH signal at the start of the plotted interval, $t_0$,
using $\Phi_0=\operatorname{atan2}\!\left(h_\times^{\mathrm{SPH}}(t_0),
h_+^{\mathrm{SPH}}(t_0)\right)$.
The analytical curves are initialized with this same phase,
so their agreement in phase at $\Delta t=0$ is imposed by
construction and is not an independent test of their phase evolution.
After the initial alignment, each description follows its adopted phase evolution.

\begin{figure}[h]
  \centering
  \includegraphics[width=\textwidth]{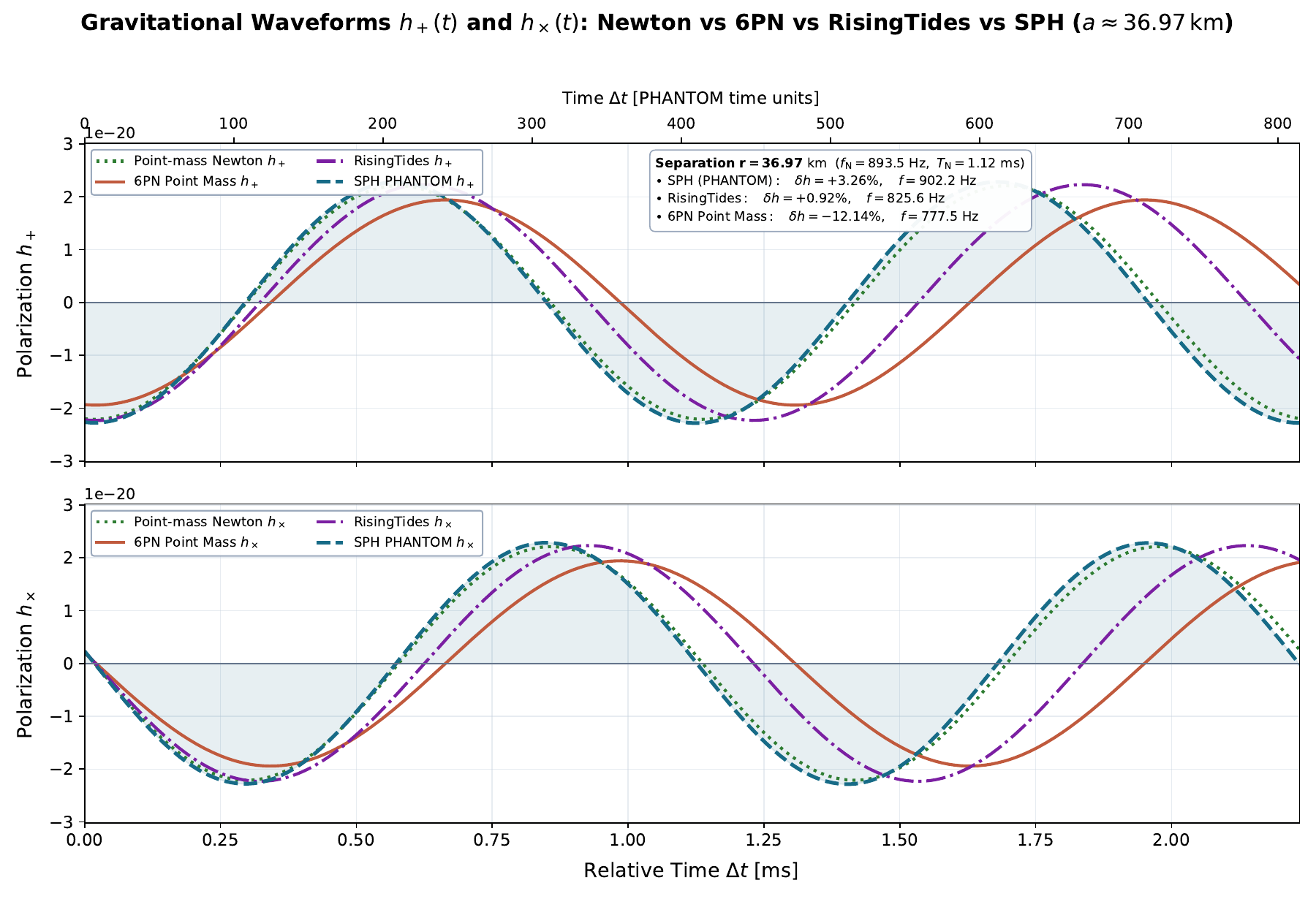}
  \caption{Face-on plus and cross polarizations for the quasi-stationary BNS
    configuration at $a\simeq36.97$~km.
    The curves share the same initial GW phase.
    The in-panel $\delta h_X$ values give the percentage amplitude deviations
  from the Newtonian result; $f_{\mathrm{GW,N}}$ is the Newtonian GW frequency.}
  \label{fig:waveform-comparison-a37km}
\end{figure}

\begin{figure}[h]
  \centering
  \includegraphics[width=\textwidth]{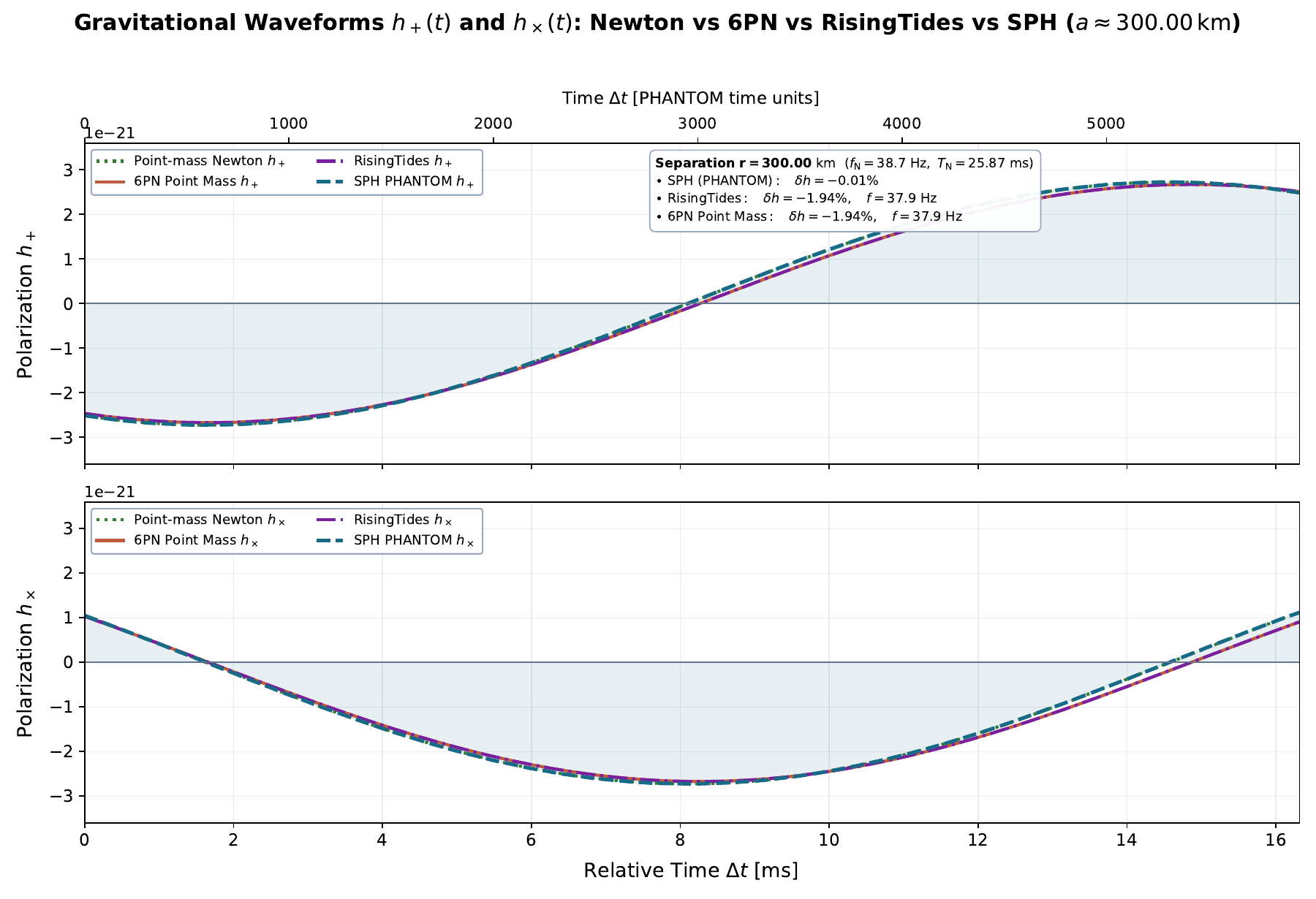}
  \caption{Face-on plus and cross polarizations for the quasi-stationary BNS
    configuration at $a=300$~km.
    The curves share the same initial GW phase.
    The in-panel $\delta h_X$ values give the percentage amplitude deviations
  from the Newtonian result; $f_{\mathrm{GW,N}}$ is the Newtonian GW frequency.}
  \label{fig:waveform-comparison-a300km}
\end{figure}

At \(a\simeq36.97\)~km (Fig.~\ref{fig:waveform-comparison-a37km}), phase
differences become visible over the plotted interval: the 6PN and \code{Rising
Tides} curves progressively separate from the Newtonian and SPH curves.
The models also differ in amplitude, as indicated by the in-panel $\delta h_X$ values.
At \(a=300\)~km (Fig.~\ref{fig:waveform-comparison-a300km}), the four predictions
remain much closer in both phase and amplitude.
The larger divergence at the smaller separation
reflects the increasing importance of relativistic and tidal effects. Since
the initial phase is aligned by construction in both figures, only the
subsequent phase evolution is being compared.

\subsection{Amplitude envelopes as functions of separation}

Based on the waveform comparison above, we now compare the strain envelopes
of the four descriptions as functions of orbital separation.
The polarizations oscillate, so we use their face-on envelope rather than the instantaneous value
of either polarization.
For a model $X$ at separation $r$, we define the face-on strain amplitude as
\begin{equation}\label{eq:strain-envelope}
  h_X(r) =
  \sqrt{\bigl[h_+^X(t;r)\bigr]^2+\bigl[h_\times^X(t;r)\bigr]^2},
  \qquad X\in\{\mathrm{N},\mathrm{6PN},\mathrm{RT},\mathrm{SPH}\}.
\end{equation}
At leading quadrupole order for a circular, face-on binary the polarizations
are 90 degrees out of phase, so this norm is independent of phase and gives the usual waveform envelope.
We use the same norm as an envelope for all four model comparisons.
For each $h_{\rm SPH}(r)$, we start from an HPR-relaxed binary and follow its
short-term evolution over several GW cycles, with the separation $r$ kept nearly
constant.
We evaluate the instantaneous face-on envelope from
Eq.~\ref{eq:strain-envelope} throughout this interval.
We use the post-HPR time averages of both the envelope and the orbital
separation as the amplitude and separation coordinate of each SPH data point.
Their temporal standard deviations are retained as the vertical and
horizontal error estimates, respectively.

\begin{figure}[h]
  \centering
  \includegraphics[width=\textwidth]{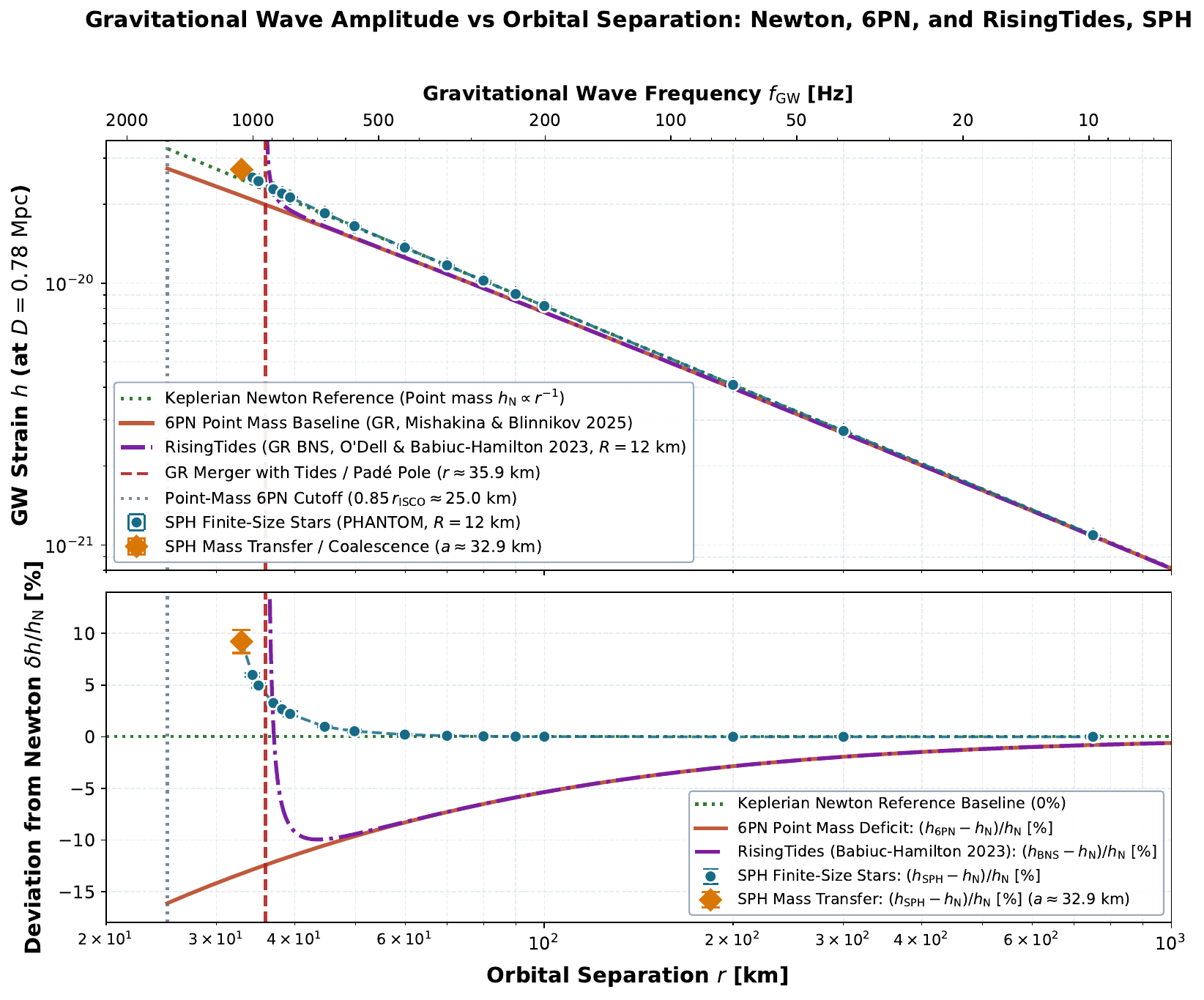}
  \caption{Strain-envelope comparison for the Newtonian point-mass,
    6PN point-mass,
    \code{Rising Tides},
    and SPH descriptions.
    The upper panel shows $h_X(r)$.
    The lower panel shows the amplitude deviation from the Newtonian
    reference in percent, $(\delta h_X/h_{\mathrm{N}})\times100\%$,
    where $\delta h_X=h_X-h_{\mathrm{N}}$.
    Blue markers denote SPH finite-size configurations, and
    the orange marker denotes the SPH mass-transfer/coalescence configuration.
    The rightmost SPH point corresponds to $f_{\mathrm{GW}}\sim10\,\mathrm{Hz}$,
    close to the lower-frequency threshold adopted for LIGO~\citep{LIGOScientificCollaboration2015}.
  }
  \label{fig:unifiedComp}
\end{figure}

Figure~\ref{fig:unifiedComp} shows the four strain amplitude envelopes
as functions of orbital separation in the upper panel
and their relative deviations from the Newtonian envelope in the lower panel.
We denote the four curves by $h_{\mathrm{N}}(r)$ for the Newtonian
point-mass reference, $h_{\mathrm{6PN}}(r)$ for the relativistic
point-mass waveform, $h_{\mathrm{RT}}(r)$ for the NR-calibrated
\code{Rising Tides} model, and $h_{\mathrm{SPH}}(r)$ for the
quadrupole extracted from the HPR-relaxed particle configurations.
The dotted vertical line marks the adopted 6PN point-mass cutoff at
$0.85\,r_{\mathrm{ISCO}}\simeq25\,\mathrm{km}$, near the innermost stable
circular orbit (ISCO).
The dashed line marks the formal pole of the Pad\'e approximant
used for the tidal-phase fit in \code{Rising Tides}~\citep{ODellHamilton2024},
at $r_{\mathrm{pole}}\simeq35.9\,\mathrm{km}$.
The orange SPH marker denotes mass transfer/coalescence at $a_{\mathrm{MT}}\simeq32.9\,\mathrm{km}$.
These markers are not directly comparable.
The Pad\'e pole is a formal singularity of the analytic fit, not a physical contact radius.

By Eqs.~\ref{eq:newtonian-waveforms} and~\ref{eq:strain-envelope}, the
Newtonian envelope is $h_{\mathrm{N}}(r)\propto r^{-1}$ and therefore
increases monotonically as the separation decreases.
In the plotted range, the 6PN and \code{Rising Tides}
envelopes also increase as the separation shrinks.
The SPH samples show the same trend, measured from the relaxed configurations.

We now examine why the SPH envelope lies above the Newtonian reference.
For a binary of two extended bodies, the leading-order source quadrupole can be
decomposed into orbital and stellar contributions as
\begin{equation}\label{eq:source-quadrupole-decomposition}
  Q_{ij}^{\mathrm{sys}}=Q_{ij}^{\mathrm{orb}}+
  \sum_{A=1}^{2}\left(Q_{Aij}^{\mathrm{spin}}+Q_{Aij}^{\mathrm{tide}}\right).
\end{equation}
Here $Q_{ij}^{\mathrm{orb}}$ is the orbital quadrupole of the two stars,
treated as point masses in the binary center-of-mass frame, while
$Q_{Aij}^{\mathrm{spin}}$ and $Q_{Aij}^{\mathrm{tide}}$ are the quadrupoles
induced by stellar rotation and the companion's tide, respectively; the latter is defined in
Eq.~\ref{eq:tidal-love-response}.
In our corotating models, each stellar spin quadrupole $Q_{Aij}^{\mathrm{spin}}$
is axisymmetric and stationary in the inertial frame,
so its second time derivative vanishes.
In the linear, large-separation limit,
$\mathcal{E}_{Aij}\sim GM_B/r^3$ (see Eq.~\ref{eq:tidal-love-response}) and
$Q_{Aij}^{\mathrm{tide}}\sim\lambda_A GM_B/r^3$, whereas
$Q_{ij}^{\mathrm{orb}}\sim\mu r^2.$
At fixed separation, the relaxed configuration is stationary in the frame
rotating with the binary.
The orbital and tide-induced quadrupoles therefore
rotate in the inertial frame with orbital phase $\Phi(t)$.
Their time-dependent components vary as $\cos 2\Phi$ and $\sin 2\Phi$.
Thus, at fixed separation, taking second time derivatives multiplies these components by
$-(2\Omega)^2$.
Since the strain is proportional to $\ddot Q_{ij}^{\mathrm{sys}}$
by Eq.~\ref{eq:quadrupole-strain},
and the time-dependent orbital and tidal
components add for our face-on configurations,
the leading-order estimate for the face-on amplitude ratio is
\[
  \begin{aligned}
    \frac{h_{\mathrm{SPH}}(r)}{h_{\mathrm{N}}(r)}
    =
    1 + \frac{\left|\sum_A\ddot Q_{Aij}^{\mathrm{tide}}\right|}{|\ddot Q_{ij}^{\mathrm{orb}}|}
    &\simeq 1
    + k_{2,A}\frac{M_B}{\mu}\left(\frac{R_A}{r}\right)^5
    + k_{2,B}\frac{M_A}{\mu}\left(\frac{R_B}{r}\right)^5
    >1.
  \end{aligned}
\]
up to order-unity projection factors. This estimate applies in the linear
regime; near mass transfer the response becomes nonlinear.

By contrast, the PN curves lie below the Newtonian reference. For equal
masses, the large-separation expansion of the 6PN amplitude in the radius
convention used here is~\citep{MishakinaBlinnikov2025a}
\[
  \frac{h_{\mathrm{6PN}}(r)}{h_{\mathrm{N}}(r)}
  =1-\frac{3-\eta}{2}\frac{GM}{c^2r}
  +\mathcal{O}\!\left(\frac{G^2M^2}{c^4r^2}\right)
  =1-1.375\frac{GM}{c^2r}
  +\mathcal{O}\!\left(\frac{G^2M^2}{c^4r^2}\right) < 1,
  \qquad \eta=\frac14.
\]

\subsection{Tidal Excess and Love-Number Rescaling}

The comparison above concerns the full amplitude envelopes, whose differences
combine finite-size effects with the relativistic correction to the point-mass
signal.
To compare the tidal response itself,
we subtract the point-mass baseline used by each model.
We refer to each resulting difference as a tidal
excess (or tidal addition).
For SPH, the relevant baseline is the Newtonian
point-mass envelope, so we define
\begin{equation}\label{eq:sph-tidal-excess}
  \Delta h_{\mathrm{SPH}}^{\mathrm{tide,N}}(r)
  \equiv h_{\mathrm{SPH}}(r)-h_{\mathrm{N}}(r).
\end{equation}
Thus, the SPH deviation in the lower panel of Fig.~\ref{fig:unifiedComp} is
this excess normalized by $h_{\mathrm{N}}$ and expressed as a percentage.
For the relativistic model, the corresponding tidal addition is measured relative
to its 6PN point-mass baseline:
\begin{equation}\label{eq:rt-tidal-excess}
  \Delta h_{\mathrm{RT}}^{\mathrm{tide}}(r)\equiv h_{\mathrm{RT}}(r)-h_{\mathrm{6PN}}(r).
\end{equation}

The quadrupole formula for the strain, together with the source-quadrupole
decomposition in Eq.~\ref{eq:source-quadrupole-decomposition} and the tidal
response in Eq.~\ref{eq:tidal-love-response}, implies the leading-order scaling
for the equal-mass BNS at fixed masses, radii, and orbital configuration:
\[
  \Delta h_{\mathrm{tide}}
  \propto \left\lVert\sum_{A=1}^{2}\ddot{Q}_{Aij}^{\mathrm{tide}}\right\rVert
  \propto \left\lVert\sum_{A=1}^{2}Q_{Aij}^{\mathrm{tide}}\right\rVert
  \propto \sum_{A=1}^{2}\lambda_A
  \propto k_{2}.
\]
This proportionality applies to the \textbf{linear} tidal response and does not describe nonlinear strong-field effects.

The SPH calculation is Newtonian and therefore uses the Newtonian
Love number $k_2^{\mathrm{N}}$,
whereas the relativistic \code{Rising Tides} model uses the GR tidal response,
characterized at leading order by $k_2^{\mathrm{GR}}$
\citep{Hinderer2008, Hinderer2009Erratum}.
We therefore rescale the fractional SPH tidal excess,
$\Delta h_{\mathrm{SPH}}^{\mathrm{tide,N}}/h_{\mathrm{N}}$, by the Love-number ratio:
\begin{equation}\label{eq:love-number-rescaling}
  \xi_{\mathrm{N}\to\mathrm{GR}}
  =\frac{k_2^{\mathrm{GR}}}{k_2^{\mathrm{N}}}
  =\frac{0.0520}{0.2599}\simeq0.200,
  \qquad
  \frac{\Delta h_{\mathrm{SPH}}^{\mathrm{tide,GR}}(r)}{h_{\mathrm{N}}(r)}
  \simeq \xi_{\mathrm{N}\to\mathrm{GR}}
  \frac{\Delta h_{\mathrm{SPH}}^{\mathrm{tide,N}}(r)}{h_{\mathrm{N}}(r)}.
\end{equation}

\begin{figure}[h]
  \centering
  \includegraphics[width=\textwidth]{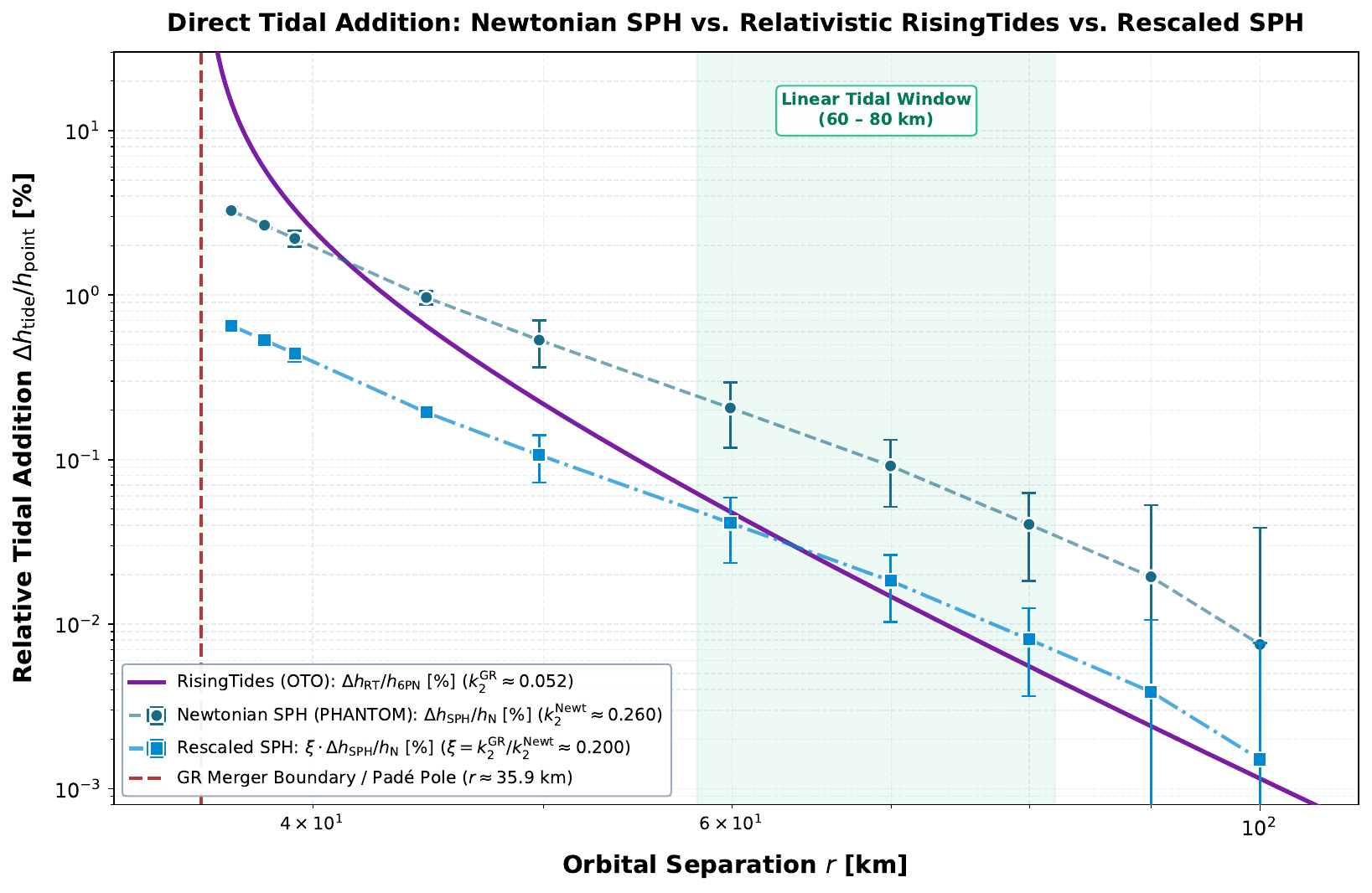}
  \caption{Relative tidal-amplitude additions versus orbital separation.
    The purple curve shows \code{Rising Tides}, the teal circles show the
    Newtonian SPH excess, and the blue squares show that excess rescaled by
    $k_2^{\mathrm{GR}}/k_2^{\mathrm{N}}$.
    SPH error bars are calculated from post-HPR time series:
    $\sigma_r=\mathrm{std}(r(t))$ horizontally and
    $y_{\mathrm{err}}=\sigma_h/h_{\mathrm{N}}(\langle r\rangle)\times100\%$
    vertically, where $\sigma_h=\mathrm{std}(h(t))$; for blue squares,
    $y_{\mathrm{err}}$ is multiplied by $\xi_{\mathrm{N}\to\mathrm{GR}}$.
    The green band marks the linear-tide interval ($60$--$80\,\mathrm{km}$).
    The dashed vertical line marks the formal Pad\'e pole, not a
  physical merger boundary.}
  \label{fig:tide-rescaling}
\end{figure}

Figure~\ref{fig:tide-rescaling} compares the fractional \code{Rising Tides}
addition, $\Delta h_{\mathrm{RT}}^{\mathrm{tide}}(r)/h_{\mathrm{6PN}}(r)$;
the Newtonian SPH tidal excess,
$\Delta h_{\mathrm{SPH}}^{\mathrm{tide,N}}(r)/h_{\mathrm{N}}(r)$; and the
Love-number-rescaled SPH excess,
$\Delta h_{\mathrm{SPH}}^{\mathrm{tide,GR}}(r)/h_{\mathrm{N}}(r)$,
all relative to their corresponding point-mass baselines and expressed in percent.
The SPH error bars are calculated from the post-HPR time series:
the horizontal error is $\sigma_r=\mathrm{std}(r(t))$,
while the vertical error is
$y_{\mathrm{err}}=\sigma_h/h_{\mathrm{N}}(\langle r\rangle)\times100\%$,
where $\sigma_h=\mathrm{std}(h(t))$; for the rescaled points,
$y_{\mathrm{err}}$ is multiplied by $\xi_{\mathrm{N}\to\mathrm{GR}}$.
The post-HPR scatter in the absolute amplitude, $\sigma_h$, remains of
similar order across the sampled separations, whereas
$h_{\mathrm{N}}\propto r^{-1}$.
Consequently, the fractional scatter
$\sigma_h/h_{\mathrm{N}}$ grows with $r$ and, by $r\simeq80\,\mathrm{km}$,
exceeds the Love-number-rescaled tidal excess,
so the tidal contribution becomes unresolved at larger separations.
Within the green band, the vertical $1\sigma$ scatter
is about $48\%$ of the SPH tidal excess on average.
At separations below the green $60$--$80\,\mathrm{km}$ band, the rescaled
SPH results depart from \code{Rising Tides}, indicating that the linear
Love-number rescaling no longer describes the nonlinear tidal response.
At the resolved points in the green band, the mean relative difference
between the rescaled SPH results and \code{Rising Tides}
is approximately $15\%$.

The SPH configurations are corotating, whereas the \code{Rising Tides}
calibration uses nonspinning binaries~\citep{ODellHamilton2024}.
If HPR could relax a nonrotating binary, we could calculate its Newtonian
fractional tidal excess directly and obtain the target GR-rescaled value as
\begin{equation}\label{eq:nonrot-gr-tidal-excess}
  \frac{\Delta h_{\mathrm{nonrot}}^{\mathrm{tide,GR}}(r)}{h_{\mathrm{N}}(r)}
  \equiv
  \frac{k_2^{\mathrm{GR}}}{k_{2,\mathrm{N}}^{\mathrm{nonrot}}}
  \frac{\Delta h_{\mathrm{nonrot}}^{\mathrm{tide,N}}(r)}{h_{\mathrm{N}}(r)},
\end{equation}
where $k_{2,\mathrm{N}}^{\mathrm{nonrot}}$ is the dynamic Love number of a
nonrotating star, characterizing its response to a periodically varying tidal
field~\citep{PnigourasEtAl2023,DewberryLai2022}.
However, the HPR procedure used here produces only corotating configurations,
so this nonrotating SPH quantity cannot be obtained directly.
Below, we estimate the magnitude of the systematic error,
$\varepsilon_{\mathrm{res}}(r)$, defined as the relative difference between
the result obtained from Eq.~(\ref{eq:love-number-rescaling}) and the target
value given by Eq.~(\ref{eq:nonrot-gr-tidal-excess}).

In the linear-tide regime, for equal masses and radii, the fractional tidal
amplitude excess is proportional to the Love number:
\begin{equation}\label{eq:tidal-excess-love-scaling}
  \frac{\Delta h_X^{\mathrm{tide,N}}}{h_{\mathrm{N}}}
  \propto k_{2,\mathrm{N}}^{X},
  \qquad X=\mathrm{corot},\,\mathrm{nonrot}.
\end{equation}
Here, $k_{2,\mathrm{N}}^{\mathrm{corot}}$ is the quasi-static Love number
describing the response of a corotating star, for which the companion's tidal
field is stationary in the corotating frame. It follows from
Eq.~(\ref{eq:nonrot-gr-tidal-excess}) and the scaling in
Eq.~(\ref{eq:tidal-excess-love-scaling}) that the dynamic Love number
$k_{2,\mathrm{N}}^{\mathrm{nonrot}}$ cancels from the target expression.
Thus, it need not be calculated separately to estimate the relative error
$\varepsilon_{\mathrm{res}}(r)$.

The rotational correction to the quasi-static Love number begins at second
order in spin~\citep{PnigourasEtAl2023}:
\begin{equation}\label{eq:corot-love-expansion}
  k_{2,\mathrm{N}}^{\mathrm{corot}}
  =k_{2,\mathrm{N}}
  \left[1+\alpha_{\mathrm{rot}}\frac{\Omega_{\mathrm{orb}}^2}{\Omega_{\mathrm{dyn}}^2}
  +O\!\left(\frac{\Omega_{\mathrm{orb}}^4}{\Omega_{\mathrm{dyn}}^4}\right)\right],
  \qquad \Omega_{\mathrm{dyn}}^2\sim\frac{GM}{R^3}.
\end{equation}
Combining Eqs.~(\ref{eq:tidal-excess-love-scaling}),
(\ref{eq:love-number-rescaling}),
(\ref{eq:nonrot-gr-tidal-excess}), and
(\ref{eq:corot-love-expansion}), we obtain
\begin{equation}\label{eq:epsilon-res-love}
  \varepsilon_{\mathrm{res}}(r)
  \simeq\frac{k_{2,\mathrm{N}}^{\mathrm{corot}}}{k_{2,\mathrm{N}}}-1
  =\alpha_{\mathrm{rot}}\frac{\Omega_{\mathrm{orb}}^2}{\Omega_{\mathrm{dyn}}^2}
  +O\!\left(\frac{\Omega_{\mathrm{orb}}^4}{\Omega_{\mathrm{dyn}}^4}\right).
\end{equation}
For equal masses, $M_1=M_2=M$, we have
$\Omega_{\mathrm{orb}}^2/\Omega_{\mathrm{dyn}}^2\simeq2(R/r)^3$.
Calculations for a rotating $n=1$ polytrope~\citep{DewberryLai2022} give
$\alpha_{\mathrm{rot}}\simeq0.86$.

For $R=12\,\mathrm{km}$ and $r=60$--$80\,\mathrm{km}$, the maximum value is
$\varepsilon_{\mathrm{res}}\simeq1.4\%$. This is about ten times smaller than
the difference between SPH and \code{Rising Tides} ($\sim15\%$), and about
thirty times smaller than the mean relative $1\sigma$ scatter of the SPH tidal
excess within the green band ($\simeq48\%$). Thus, the systematic error due to
corotation is substantially smaller than the statistical scatter in the tidal
excess.

\section{Conclusions}

We have demonstrated how a Newtonian SPH code can be used to estimate the tidal contribution
to the GW amplitude at selected fixed orbital separations.

Using a BNS system with typical masses of 1.5 + 1.5 $M_\odot$ as an example --- and
by comparing our results with waveforms calibrated against full numerical GR simulations
--- we show that the tidal contribution to the GW amplitude can be estimated using as few as $\sim 10^4$ smoothed particles.
In this approach, it is necessary to apply a correction to account for the difference
in neutron star compactness between Newtonian gravity and general relativity.
Using the BNS case as an example,
the amplitude $h$ is compared across four approaches:
6PN (6th-order post-Newtonian),
\code{Rising Tides} approximation,
N (Newtonian, no tides),
and SPH (smoothed-particle hydrodynamics).
In this comparison, the deviations (6PN vs. \code{Rising Tides} and N vs. SPH) scale
according to the Love numbers $k_2$, provided that the radius $R$ is kept constant.
Applying a renormalization factor $k_2^{GR}/k_2^N$ for separations greater than 60~km yields
similar relative deviations.
At distances ($> 80$ km), the influence of tidal effects is negligible and of little practical interest.
For these separations, however, one can rely on the point-mass 6PN approximation for waveforms aimed at early detection of BNS gravitational-wave signals.

Although the simulations presented here use equal-mass neutron stars,
the method can be further developed for unequal-mass binaries.
The GR Love number depends on compactness,
so stars of different masses can have different
$k_2^{\mathrm{GR}}$ even for the same equation of
state~\citep{Hinderer2008, Hinderer2009Erratum}.
For illustration, we use an equal-mass BNS.
For unequal-mass binaries, the single Love-number rescaling coefficient
should be replaced by a weighted mean
of the two stars’ individual coefficients,
with weights given by their respective
contributions to the Newtonian tidal excess.

In this study, SPH is used only for the amplitude at fixed separations,
an HPR sequence could in principle also be extended to estimate the tidal phase correction.
This sequence could provide the quasi-static tidal quadrupole as a function of separation,
$Q_{\mathrm{tide}}(r)$.
This response could be supplied to a post-Newtonian orbital model,
where it would inform the tidal contributions to both the conservative orbital dynamics
and the time-dependent source quadrupole used to calculate GW emission.
With these contributions treated consistently,
evolving $r(t)$ and $\Omega(t)$ and integrating $\dot{\phi}=\Omega$ would yield the accumulated phase
within the adopted PN and quasi-static approximations.
We leave construction of such an orbital model to future work.

A second interesting candidate for SPH amplitude estimates is a white-dwarf--neutron-star (WDNS) binary
with typical masses of $M_{\mathrm{WD}}=0.6\,M_\odot$ and $M_{\mathrm{NS}}=1.5\,M_\odot$.
In \code{phantom}, the neutron star can be modeled as a ``sink'' particle.
Well outside Roche-lobe overflow, the white dwarf's low compactness makes a Newtonian stellar model appropriate,
and its static Love numbers are expected to satisfy $k_2^{\mathrm{N}}\approx k_2^{\mathrm{GR}}$.
A well-resolved HPR model of the white dwarf would require substantially more SPH particles than the $\sim 10^4$ used for the BNS case.
As the binary approaches Roche-lobe overflow, tidal mass transfer and disruption of the white dwarf begin quite rapidly.
Our SPH estimates indicate that Roche-lobe contact
begins much earlier than predicted by the point-mass Roche-lobe approximation of
\citet{Eggleton1983}.
Near Roche-lobe contact, the high orbital frequency and potentially asynchronous spin
can also produce time-dependent forcing in the white dwarf
and excite dynamical gravity-mode tides.
A synchronized, quasi-static HPR sequence would not capture this response~\citep{FullerLai2012}.
Dynamical tides also occur in neutron stars, but become more relevant later in the inspiral,
closer to merger~\citep{AbacEtAl2024}.
These challenges can be overcome, but other purely Newtonian effects hinder yielding useful results
with the approach that was found useful for BNS.
It remains crucial to perform a hydrostatic/hydrodynamic relaxation (HPR) of the initial models; otherwise,
the entire GW signal is lost within the numerical noise.
However, physical GW signals from internal oscillations and Newtonian deviations of the center-of-mass motion from the point-mass approximation prevent us from relying
on a simple addition of tidal perturbations to the GW waveform of point masses,
since all these effects are of the same order.

Although HPR-based Newtonian SPH does not predict relativistic orbital evolution or the accumulated GW phase,
and its tidal signal is measurable only where it exceeds the SPH noise,
it can still serve as a rapid, low-cost diagnostic of tidal-amplitude prescriptions.
At selected separations, it provides an independent estimate of the quasi-static tidal excess
and shows where that excess is distinguishable from numerical scatter and where
a point-mass amplitude remains adequate.
Although it cannot validate a complete waveform bank or replace NR simulations,
the method can help identify possible inconsistencies in GW amplitude predictions
before more expensive full general relativistic calculations are performed.

\acknowledgments{This work is supported by RSF grant 25-22-00293.
We are grateful to Andrei Yudin and Dmitriy Kudryashov for numerous critical
discussions and help.}

\bibliographystyle{mnras} 
\bibliography{refs}  

\end{document}